\documentclass[%
 reprint,
 amsmath,amssymb,
 aps,
]{revtex4-1}
\usepackage{color}
\usepackage{graphicx}
\usepackage{dcolumn}
\usepackage{bm}
\usepackage{bbold}

\begin{document}

\preprint{APS/123-QED}

\title{Complex field reversal dynamics in nanomagnetic systems}

\author{Michael Saccone}
\affiliation{%
Los Alamos National Laboratory, Los Alamos, New Mexico 87545, USA
}%

\author{Francesco Caravelli}
\affiliation{%
Los Alamos National Laboratory, Los Alamos, New Mexico 87545, USA
}%

\date{\today}

\begin{abstract}
Nanomagnetic materials, built from thin, patterned films of ferromagnetic materials, began as analogues to frustrated magnetism. Their low energy of operation and emergent properties make them strong candidates for physics based devices. A recent model of how nanomagnetic domains flip, the Glauber mean-field model, is used here to understand how systems of nanomagnets evolve when opposed by external field. This reversal can be expressed in an analytical form in the case of one-dimensional chains and trees at zero temperature, where the cascade of spin flips gives rise to harmonic power spectra. The same cascades in two and three dimensions form fractal field reversal clusters whose shape depends on the strength of the field and the tuning of interactions between nanomagnets.
\end{abstract}

\maketitle


\section{\label{sec:level1} Introduction}
To engineer physics that goes beyond that of constituent materials, scientists manufacture metamaterials using traditional and nanofabrication techniques~\cite{zheludev2012metamaterials,kadic20193d}. Some metamaterials pursue a specific functionality impossible in pure materials, such as subwavelength focusing and invisibility cloaking~\cite{hess2012active}, while others seek to emulate other physical systems with greater control over the parameter space and improved imaging of the system microstates~\cite{schiffer2021artificial, skjaervo2020advances}. More finds new ways to be different and the gestalt behaviors of these systems do not perfectly emulate their supposed counterparts due to disorder~\cite{budrikis2012disorder}, dynamical differences~\cite{morley2017vogel}, or entirely new degrees of freedom~\cite{gartside2022reconfigurable} that prove non-negligible. 

\begin{figure*}
\includegraphics[width = 1\textwidth]{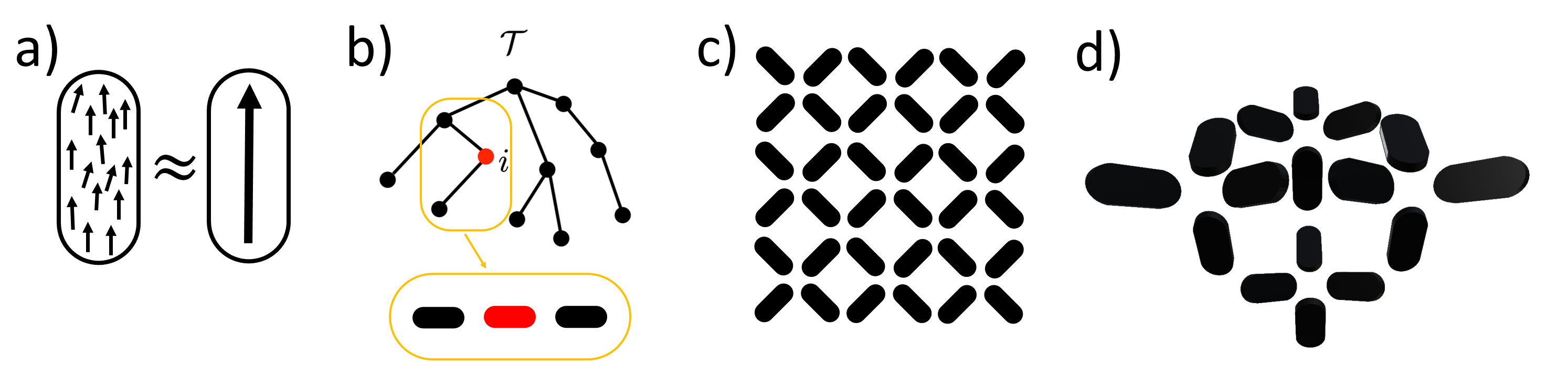}
\caption{\label{fig:lattices}a) The mean-field model sums over the thermal ensemble of spins in a nanomagnetic domain and represents their average moment as a single variable between -1 and 1, $m_i(t)$. b) An illustration of a tree of coupled nanomagnets. c) The square lattice, a two-dimensional model system for frustrated magnetism. In the nearest neighbor approximation, nanomagnets sharing a vertex interact with one another. d) A unit cell of the diamond lattice, a three-dimensional system where interaction occurs between tetrapod-sharing nanomagnets.}
\end{figure*}

In particular, the field of artificial spin  ice~\cite{schiffer2021artificial, skjaervo2020advances} has begun to grapple with this concept. Initially a means of directly imaging patterned, Ising-like nanomagnets with dipolar interactions that map onto problems in statistical physics and frustrated magnetism, the field has since grown to encompass device-oriented approaches to computation and evaluate the collective behavior of nanomagnets beyond simple, Ising spins~\cite{dion2022observation,gartside2022reconfigurable,vidamour2022reservoir}. Visualizing the nanomagnets in real time revealed that their fluctuations do not purely correspond to a thermal ensemble, but rather incorporate the complexities of relaxation pathways~\cite{arava2019engineering,farhan2013direct}, system topology~\cite{lao2018classical}, deviation from ergodicity~\cite{lammert2010direct,lao2018classical}, and innate material properties~\cite{morley2017vogel,drisko2015fepd}. While this may be modeled by micromagnetic simulations of the LLG equations~\cite{velo2020micromagnetic}, this computational approach is prohibitively costly for large systems and does not provide an analytically tractable means of understanding nanomagnet behavior. A recently introduced model of mean-field Glauber dynamics~\cite{saccone2023vertices} represents the major magnetic domains in a nanomagnetic system as mean-field, continuous variables between -1 and 1 (Fig.~\ref{fig:lattices}a) and parameterizes the interaction strength between domains. This reduces the spatial resolution of the full micromagnetic simulation for an assumed set of relevant variables. While this hides details such as the curving of magnetization at the tips of the nanomagnets, the computational overhead is significantly lowered and the increased symmetry admits insightful analytical solutions. Within this work, we leverage this model to explain why nanomagnetic systems deviate from simple Ising systems and predict what untapped dynamics should emerge under the correct conditions. Pursuant to prior experimental studies~\cite{bingham2021experimental,hallen2022dynamical,zeissler2016low,hugli2012artificial}, we begin with avalanches of spin flips in one-dimensional chains and trees (Fig.~\ref{fig:lattices}b), then expand to the more realistic two-dimensional square (Fig.~\ref{fig:lattices}c) and diamond lattice (Fig.~\ref{fig:lattices}d) systems to find complex fractal avalanches, similar to previous numerical experiments~\cite{chern2014avalanches}, and harmonic spectra. Avalanche dynamics have long been considered crucial to information processing in the brain~\cite{shriki2013neuronal} and helpful in generating neuromorphic computation in nanostructed networks~\cite{mallinson2019avalanches,pike2020atomic}. It is likely that spin ice avalanches can similarly enhance information processing.

\section{Nanomagnet chain and trees}
The model explored in this paper is that of mean-field Glauber dynamics. As derived previously for spin-ice~\cite{saccone2023vertices} as a generalization of the Contucci model~\cite{contucci2007modeling,krapivsky2010kinetic}, each nanomagnetic island is represented by a mean-field spin, $m_i \in [-1, 1]$, that evolves over time as governed by
\begin{equation}
    \dot{\vec{m}} =-\vec{m}+ \tanh (\beta Q\vec{m}+\beta \vec{h}(t)).
\end{equation}
The magnets are subject to an interaction matrix $Q$ and external field $\vec{h}(t)$, both modulated by an inverse temperature parameter $\beta$. In this deunitized result, the natural relaxation time of a domain is 1. While previous work has determined the attractors of this model and their stability for small systems\cite{saccone2023vertices}, here we expand the analysis to extensive systems with a focus on their response to the external field.

In the simplest extensive system, a one-dimensional chain, the elements of the interaction matrix are
\begin{equation}
    Q_{ij} = q \delta_{ij} + J (\delta_{i,j+1} + \delta_{i,j-1}),
\end{equation}
where $q$ is the self-energy of the system which penalizes deviation from $m_i = \pm 1$ and $J$ is the ferromagnetic (+) or antiferromagnetic (-) interaction strength of the neighboring magnets. A chain is a special case of a tree (Fig.~\ref{fig:lattices}b) with consistent degree 2, and many of the statements made about magnetization dynamics on a physical chain of nanomagnets can be extended to general trees. By approximating the governing differential equations in different regimes, we can gain insight into how the system responds to external field. For example, the high temperature or paramagnetic case linearizes the equations of motion and results in a system equivalent to driven RL circuits (See Supplementary Information).  

\subsection{Low temperature limit}
Low-temperature systems approach fully magnetized states that only change when an island's coercivity is exceeded by all fields acting on the island. This is reflected in the hyperbolic tangent function approaching a sign function, $\text{sgn}(x)$: 
\begin{equation}
    \dot{\vec{m}} =-\vec{m}+ \text{sgn}(Q\vec{m}+ \vec{h}).
\end{equation}
Changes in the moments occur in piecewise functions determined by the argument of the sign function changing from positive to negative, or vice versa. The attractors of the individual magnets are the binary values $\pm 1$. The duration of the piecewise periods may be generally difficult to determine but is solvable for simple scenarios such as a tree interaction structure (See Supplementary Information). The collective behavior of the system depends on what sequence of flips occur and how they influence one another. The initial condition of the system and the spatial pattern of the field all can evoke different behaviors. One commonly encountered scenario is a magnetization reversal from saturation, all magnets aligned in one direction, under a uniform external field. We will explore how systems behave in this deceptively simple scenario for the remainder of the paper.

For a nanomagnet chain, assume that all spins begin pointing to the left, the system is ferromagnetic $J > 0$, and that the system is subject to a static field pointing to the right, $h > 0$. The field is great enough to overcome the coercivity of a single spin, but not great enough to flip all spins at once considering their local field from ferromagnetic interactions is $2J$ when they all point to the left. Thus, $q < h < 2J$, or for a scaled field $h^* = (h - q)/2J$, $0 < h^* < 1$. Let a single nanomagnet at location $i$ start at time zero with a moment $m_i(0) = 1$ to nucleate an avalanche. This is an idealized case of a single nanomagnet flipping before its neighbors due to disorder in fabrication~\cite{budrikis2012disorder}. This magnet reverses its field on the moments at $i\pm 1$. Those nanomagnets flip their neighbors at $i\pm2$, which continue the chain until the flips terminate. Each $m$ increases monotonically once it is associated sign function potential changes sign as all arguments to the sign function only increase in magnitude throughout this process, bringing all moments into alignment with the field and their neighbors. This logic holds on a general tree as well, with magnetization flips propagating from an initial site outwards with fixed time delays determined by the strength of the field (See Supplementary Information for the full derivation). Specifically, we find that the time between moment flips for the uniform field is 
\begin{equation}
\tau = -\log (h^*).
\end{equation}
The delay is consistent for a given field, meaning that the strength of the signal neither grows nor decays, precisely the "branching fraction" of 1 determined to be optimal in neuron avalanches~\cite{shriki2013neuronal}. This time delay slowly grows infinite as the field shrinks to the threshold of $h^* = 0$ and becomes zero as the field approaches $h^* = 1$, implying infinite speed as the slightest movement of a neighboring magnet will initiate a flip. It is worth noting that this is a form of superluminal motion, in which successive events propagate at a rate that appears faster than the speed of light to an observer \cite{ginzburg1960certain}. This apparent superluminal motion is due to the fact that the external field acts on all the spins effectively independently, and $h^*$ corresponds exactly to the value at which the flipping stops being cooperative. The precise details of delayed vector potentials between nanomagnets are not accounted for and may become relevant if precise experiments near $h^* = 1$ are conducted.

If the first flip occurs at site $j$ during $t = 0$, the string propagates $R$ sites away from the initial island, and the first flipped moment is ignored due to its lack of dynamic contribution, the total magnetization of the chain is
\begin{equation}
    M(t) =2\sum_{k=0}^{R}  \big(1 - 2e^{-(t -k\tau)}\big)\Theta(t - k\tau) - \Theta(k\tau - t)
\end{equation}
where $\Theta(x)$ is the Heaviside function. For a tree with uniform degree $d$ (a Cayley tree), the number of flipping magnets grows exponentially with each layer flipped from the root moment. The magnetization is then
\begin{eqnarray}
    M(t) &=& d\big[(1 - 2e^{-t}\big)\Theta(t) - \Theta(-t) \big]\nonumber \\
    &+& d\sum_{k=1}^{R}(d-1)^{k}\big(1 - 2e^{-(t -k\tau)}\big)\Theta(t - k\tau)\nonumber \\
    &&- \Theta(k\tau - t).
\end{eqnarray}




To connect to studies of magnetic noise spectra~\cite{dusad2019magnetic} and generally elucidate the time-series, we look to the power spectrum, $P(\omega) = |\mathcal{F}(M(t))|.^2$, where $\mathcal{F}(x(t))$ is the Fourier transform of a function. For the chain, this gives
\begin{equation}
    P(\omega) =  \frac{8}{\pi} \frac{1}{\omega^4 + \omega^2} \frac{1 - \cos R\tau\omega}{1 - \cos \tau\omega}.
\end{equation}

\begin{figure*}
\includegraphics[width = .9\textwidth]{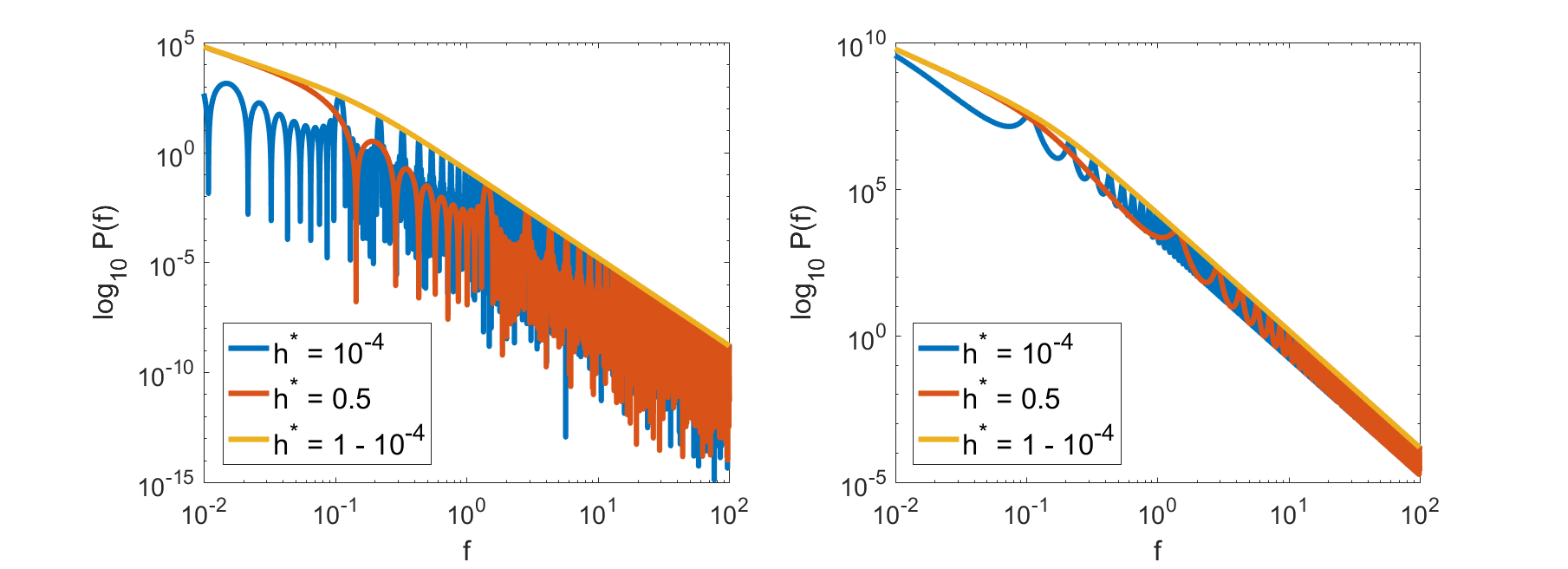}
\caption{\label{fig:stringSpectra}The power spectra $P(f)$ for $R = 10$ plotted as the external field $h^*$ is varied for a (\textit{a}) chain of flips and (\textit{b}) a Cayley tree with $d = 3$.}
\end{figure*}
Holistically, the power spectrum is enveloped by $\frac{1}{(\omega^4 + \omega^2)}$, quickly approaching $\frac{1}{\omega^4}$, which is unitless because the natural relaxation frequency of a single nanomagnet is set to 1 in the governing differential equations. A signal like this would be hidden in magnetization noise due to the $1/\omega^4$ envelope, masked by the stronger $1/\omega^2$ signals from Brownian motion at higher temperatures or stronger still subdiffusive motion of monopoles at lower temperatures~\cite{nisoli2021color}. The driven motion of the magnetization motion has the same $1/\omega^4$ signature as superdiffusivity~\cite{nisoli2021color}. The rest of the function is modulated by time constants $\tau$ and $R\tau$, giving it a harmonic structure that is audible when converted to the audio output (see Supplementary Material) and visible when the spectra are plotted (Fig.~\ref{fig:stringSpectra}). The positions of these harmonics depend on the strength of the external field, as seen in Fig.~\ref{fig:stringSpectra}a), and vary most rapidly around $h^* = 1$, the transition to collective spin flipping. Increasing the number of spin flips adds more ripples between primary harmonics. The form of the Cayley tree power for $d = 3$ is similar outside of more complex harmonic terms:
\begin{equation}
    P(\omega) =  \frac{18}{\pi} \frac{1}{\omega^4 + \omega^2} \frac{1 + 4^{R+1} - 2^{R+2}\cos (R+1)\tau\omega }{5 - 4\cos \tau\omega}.
\end{equation}
The scaling envelope is the same, but the power is much larger due to the exponential growth with increasing $R$, while the harmonics are proportionally weaker (Fig.~\ref{fig:stringSpectra}b)

Note that multiple of these processes may occur at the same or staggered times and they will be no different aside from the early collision of domain walls with one another. The specific site chosen to begin the propagation may occur experimentally due to variance in the coercivity of individual nanomagnets or spatial fluctuation of the field. When the field is increased gradually, this is more reasonable to expect than the simultaneous switching of all nanomagnets. As will be discussed more in the conclusion, this also explains why experiments show chains of nanomagnets flipping in rapid succession despite system dynamics still occurring on a longer timescale. Hypothetically, a thermal flip may encourage a series of "kinetic" flips, shown here to require no addition of energy from the environment, which then terminate due to a higher local coercivity.

\begin{figure*}
\includegraphics[width = .95\textwidth]{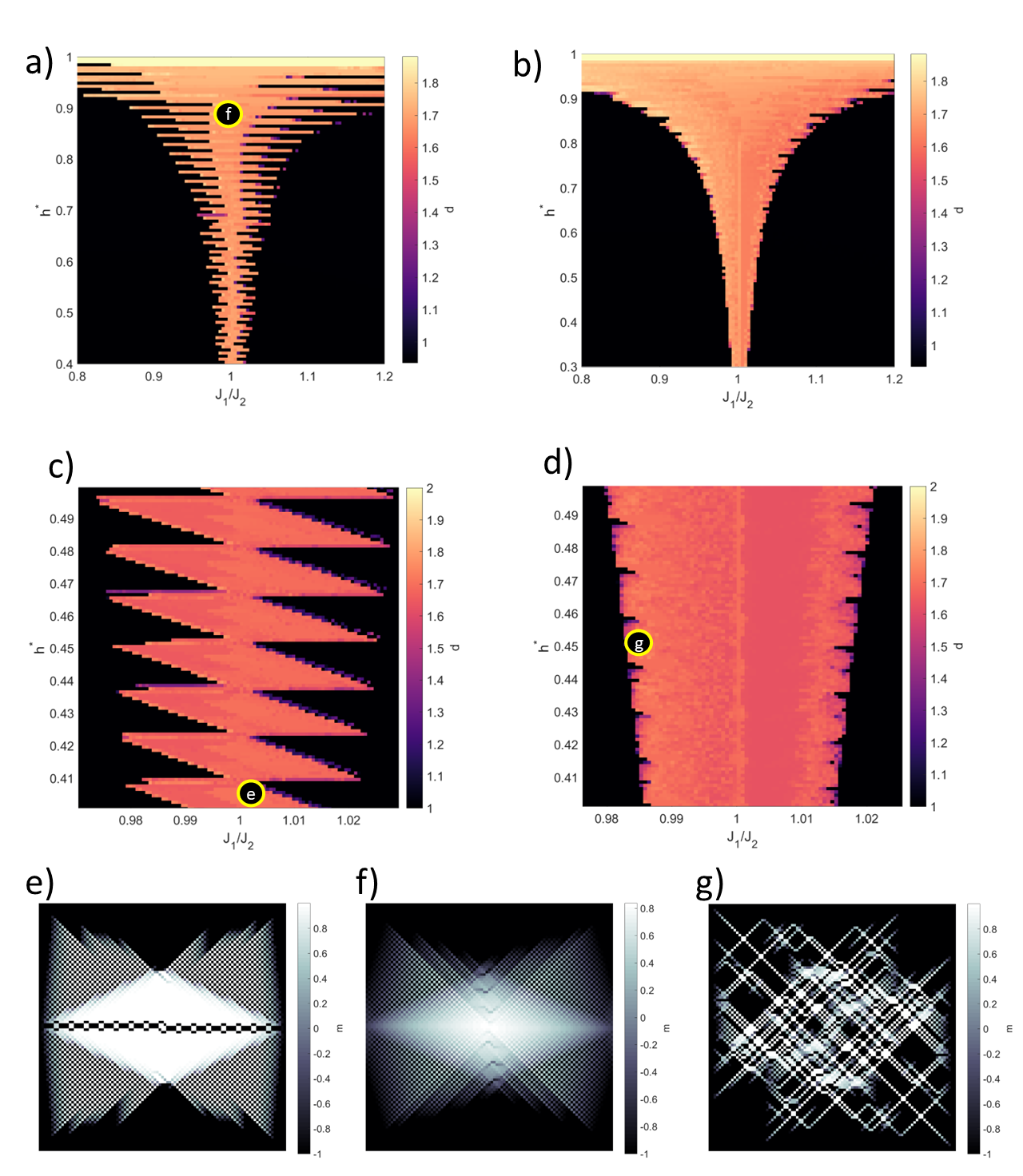}
\caption{\label{fig:squarePhase}The fractal dimension of the field reversal clusters at a) $\beta = 2000$ and b) $\beta = 200$ plotted against interaction strength ratio $J_1/J_2$ and scaled field $h^*$. Increasing the resolution of the diagram reveals that the jagged structure of the diagram in c) rapidly becomes smoother as temperature increases in d). e) At low temperature ($\beta = 2000$), intermediate external field ($h^* = 0.40$), and equivalent interaction strengths ($J_1/J_2 = 1$), each spin's magnetic moment $m$ in the field reversal simulation is plotted as a heatmap. Considering the cluster to be all moments greater than 0, the fractal dimension of this shape is 1.71. f) When the field is increased to $h^* = 0.90$ ceteris paribus, the fractal dimension remains the same while moments tend to flip at lower values of their neighboring moments. g) At an increased temperature ($\beta = 200$), intermediate field ($h^* = 0.45$), and on the edge of the interaction strength ratio that results in fractal domains ($J_1/J_2 = 0.9827$), the moments tend to reverse in single strings but occasionally bifurcate into more fractious structures, resulting in a fractal dimension of 1.65.}
\end{figure*}

\section{Square Lattice}
A particularly captivating feature of spin ice is vertex frustration: the symmetry of interaction energies where spins meet at a vertex means they may take on several orientations while maintaining the same local energy. Artificial square ice connects vertices on a square grid with nanomagnetic edges~\cite{moller2006artificial}. The ground state, when all interactions are symmetric and local, arranges spins into two-in/two-out configurations (Fig.~\ref{fig:lattices}c). The magnets in this pattern may be labelled by $i_x$ and $i_y$, each with the range $[0,\sqrt{N} - 1]$, that correspond to $i = i_x + \sqrt{N} i_y$, making the interaction matrix
\begin{multline}
    Q_{ij} = q \delta_{ij} + J_1 (\delta_{i_x,j_x + 1}\delta_{i_y,j_y} + \delta_{i_x,j_x - 1}\delta_{i_y,j_y}- \delta_{i_x,j_x}\delta_{i_y,j_y+1} \\-  \delta_{i_x,j_x}\delta_{i_y,j_y-1}) + J_2(\delta_{i_x,j_x-1}\delta_{i_y,j_y-1}+  \delta_{i_x,j_x+1}\delta_{i_y,j_y+1}),
\end{multline}
when $i_x + i_y$ is even and 
\begin{multline}
    Q_{ij} = q \delta_{ij} + J_1 (\delta_{i_x,j_x + 1}\delta_{i_y,j_y} + \delta_{i_x,j_x - 1}\delta_{i_y,j_y}- \delta_{i_x,j_x}\delta_{i_y,j_y+1} \\-  \delta_{i_x,j_x}\delta_{i_y,j_y-1}) + J_2(\delta_{i_x,j_x+1}\delta_{i_y,j_y-1}+  \delta_{i_x,j_x-1}\delta_{i_y,j_y+1}),
\end{multline}
when $i_x + i_y$ is odd. $J_1$ is the typically stronger interaction energy of perpendicular magnets and $J_2$ is the typically weaker energy of collinear magnets. Various experiments use fabrication techniques to modify the ratio of $J_1/J_2$, typically reducing $J_1$ to restore symmetry to the model~\cite{perrin2016extensive,ostman2018interaction,farhan2019emergent}. Despite the additional interactions, several modes of the system's field response are largely the same. In the high-temperature limit, the homogeneous solution decays and the particular solution still responds site by site to the external field. All previous conclusions about field response being RL circuit-like hold. The collective modes in the low-temperature approximation also do not occur below $h = q$ and flip the whole system simultaneously above $h = q + 2J$, encouraging the use of the same scaled field $h^*$. 

The most substantial new response lies in sequential spin flips. There are parameter regimes in which a single string of spins flips, but the underlying spin texture and types of allowed transitions influence the sequence of flips. We simulate the field reversal of a saturated square ice with periodic boundary  conditions and a single, central nanomagnet with no coercivity to seed a domain flip. Many transitions not anticipated through single spin flips occur, such as direct type I to type II conversion when the delay between two flips is low and the ratio of $J_1/J_2$ is close to 1. The process cannot be simplified to evenly delayed spin flips but the cluster grows at its edge as new transitions become possible. Flipped spins beget more flipped spins, analogous to diffusion-limited aggregation (DLA)~\cite{daccord1986radial,mathiesen2006universality}. DLA is a process of cluster growth in which randomly walking particles are attached to an existing cluster when they contact it. A relatively simple ruleset results in fractal structures with characteristic fractal dimensions around 1.7~\cite{daccord1986radial,mathiesen2006universality} when embedded in two dimensions and around 2.5 when embedded in three dimensions~\cite{tokuyama1984fractal}. That is, the fractal cluster only occupies a fraction of the whole space in which it exists, but is not constrained to an integer lower dimension such as a line or plane. To compare magnetic reversal clusters to DLA, the fractal nature was quantified with a network approach to fractal dimension. This dimension compares the accumulation of points on subsets of the network, $m$, which are all a distance $r$ traversed from arbitrarily chosen central points. The growth rate is assumed to be a power law, as is the case with area or volume, $m = Ar^d$, where $A$ is the rate of scaling and $d$ is the dimension of the cluster. $d$ is then extracted from the standard least squares fit of the relationship between $m$ and $r$.

\begin{figure*}
\includegraphics[width = 1\textwidth]{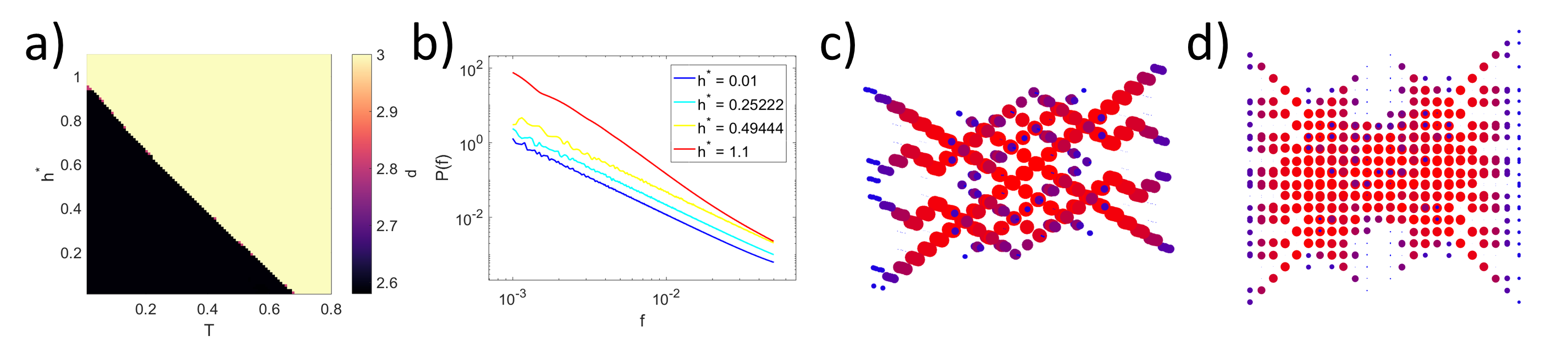}
\caption{\label{fig:pyroAnalysis} a) The fractal dimension of the field reversal cluster on a pyrochlore lattice plotted versus temperature $T$ and scaled field $h^*$. b) The power spectra of the magnetization time-series at zero temperature for increasing values of the field. c) Visualization of the moments on the pyrochlore lattice after field reversal with $h^* = 0.1$ and $T = 0$ as seen from slightly off the (011) direction and the d) (001) direction. The area of the dots scales with the deviation of the moments from their initial value of -1 and the color linearly scales between blue and red as the moments reverse.}
\end{figure*}
In Fig.~\ref{fig:squarePhase}(a-b) we plot a heatmap of the fractal dimension against the ratio of $J_1/J_2$ and the external field strength at $\beta = 2000$ (Fig.~\ref{fig:squarePhase}a) and 200 (Fig.~\ref{fig:squarePhase}a). A complex phase diagram is revealed, showing distinct regions where $f = $ 1, 2, and values between 1.6 to 1.8. Though the precise transitions that govern the aggregation of a fractal domain vary wildly throughout the phase space and accordingly change the exact value of the fractal dimension (see Supplementary Material), the non-integer dimensions observed correspond to additional work on diffusion-limited aggregation which can fluctuate based on the precise assumptions of the simulation or experimental scenarios. One-dimensional reversal occurs along chains of collinear moments when $J_1/J_2$ is less than a critical, field-dependent value, and along perpendicular moments when $J_1/J_2$ is above another value, the direction of propagation reflects stronger interactions. For finite temperatures, there is a region of high $h^*$ less than 1 where all spins flip simultaneously, making a cluster of dimension 2. The exact shape of the boundary between these phases is detailed and, as zooming in on the diagram reveals in Fig.~\ref{fig:squarePhase}c-d), loses detail as the temperature is increased. The polygonal structure at $\beta = 2000$ (Fig.~\ref{fig:squarePhase}c) rapidly becomes smoother as temperature increases to $\beta = 200$ (Fig.~\ref{fig:squarePhase}d). Although the details of this phase may be difficult to observe experimentally, the diagram suggests a stark transition between one-dimensional reversal and fractal reversal, then fractal to total reversal at critical fields when $J_1 \neq J_2$, simply observable with precise field control. The variety of shapes observable is broad, particularly near the phase boundary. While interaction energies are equivalent ($J_1/J_2 = 1$), the temperature is relatively low ($\beta = 2000$), and the field is moderate ($h^* = 0.40$), the cluster is highly symmetric with a fractal dimension of $f = 1.73$ (Fig.~\ref{fig:squarePhase}e). Increasing the field to $h^* = 0.90$ (Fig.~\ref{fig:squarePhase}f) makes the moments flip earlier in the motion of their neighbors, resulting in a similar shape and fractal dimension of $f = 1.71$. At a higher temperature ($\beta = 200$) and asymmetric interaction energies ($J_1/J_2 = 0.9827$), the cluster forms through strings of reversals that occasionally bifurcate into branching structures before recombining again into strings (Fig.~\ref{fig:squarePhase}g). This reduces the fractal dimension to 1.65. Altogether, the strength of the field, temperature, and balance of the interaction strengths can alter the fractal dimension and holistic appearance of fractal domains.

\section{Diamond Lattice}

Geometric frustration naturally occurs on the diamond lattice~\cite{ramirez1999zero}, a three-dimensional structure of connected tetrapods. When the legs of the tetrapods are interacting units, such as hydrogen atoms in water ice or exchange coupled electron orbitals with nonzero spin in magnetic pyrochlore materials, the symmetry of the interactions on the tetrapods gives rise to ice physics via the energy equivalence of 6 "two-in, two-out" configurations. This inspired the two-dimensional equivalent, the square lattice, but unless perturbed the diamond lattice does not admit the same asymmetry between $J_1$ and $J_2$ interaction energies. Recent interest in pure magnetic pyrochlore crystals explores the spectra of their magnetic noise as a function of the applied field and found that fractal structures of moment reversal mediated by the path of effective monopoles explains features of the noise~\cite{hallen2022dynamical}. Though the model presented here is not valid for single electron spins, the noise profile of artificial spin ice has recently been explored as a function of field strength and direction~\cite{goryca2022magnetic} and nanomagnetic pyrochlore systems have recently been constructed~\cite{may2021magnetic} and show promising potential to image their moments under field reversal.

We perform another set of field reversal experiments on a pyrochlore system with symmetric nearest-neighbor interactions to explore what structures emerge from the magnetic moments and the spectra of the magnetization time-series. Once more, a central, nucleating spin is set to $m_i(0) = 1$. The scaled field, $h^* = (h - q)/2J$, is ran between 0 and 1.1 and applied in the (100) direction, opposite the initial magnetization in the (-100) direction. Just after $h^* = 0$ the moments adjacent to the nucleating moment can begin to reverse at all. A unit cell of 16 moments is tiled in a 5x5x5 cubic grid with periodic boundary conditions for a total of 2000 moments. Higher than $h^* = 1$ and all moments collectively reverse, immediately overcoming their local interactions. As temperature increases, so does the potential to collectively reverse via the softening of the moments. To characterize the shape of the reversal cluster, its fractal dimension is calculated. As observed in Fig.~\ref{fig:pyroAnalysis}a), there is a stark cutoff between regions in the $h^*$ and $T$ space where the fractal dimension is 2.63 and 3. A dimension of 3 corresponds to collective reversal: the whole space is filled with flipped moments. Those with a dimension of 2.63 are similar in structure to shapes formed by diffusion-limited aggregation in three dimensions, characteristically having a dimension of around 2.5~\cite{tokuyama1984fractal}. Once more, the process here is distinct from diffusion-limited aggregation in that it is entirely deterministic, but the similarity in fractal dimension suggests parallels in their physics. The cluster grows by following a complex series of corridors that are forbidden or permitted based on the initial state of unchanged, saturated moments and the evolving moments from the growing cluster. The time-series of the magnetic moment does not produce harmonics due to less consistent time delay but only drops off with a $1/\omega^2$ scaling below $h^* = 1$ (Fig~\ref{fig:pyroAnalysis}b). The scaling jumps to $1/\omega^4$ above $h^* = 1$ as the system reversal is a single event. The structures of the field reversal clusters are less diverse than those of the square lattice reversals. As visualized at $h^* = 0.1$ and $T = 0$ in Fig~\ref{fig:pyroAnalysis}c-d), the fractal cluster follows largely the same path as field and temperature are increased, until the point where the entire system flips all at once.

\section{Conclusion}
The adaptation of nanomagnets for device applications~\cite{arava2019engineering,gartside2022reconfigurable,saccone2022direct} requires a better understanding of collective out-of-equilibrium dynamics. The dynamical modes modeled here confirm one-dimensional moment reversal avalanches~\cite{bingham2021experimental} can arise from the Glauber mean-field dynamics and produce harmonic, superdiffusive spectra. The avalanches are stable in that their velocity is constant, an ideal characteristic for information processing~\cite{shriki2013neuronal}. The frequency of the harmonics is tunable by the strength of the external field, making the system sensitive to different strengths of inputs. When interaction strengths are near equal in the square lattice and inherently for the pyrochlore lattice, the reversal clusters become fractals. Fractals like these were seen in previous simulations of artificial spin ice field reversal~\cite{chern2014avalanches}, but field reversal experiments only produced one-dimensional avalanches. Referring to Figure 3a-b, our results demonstrate that the range of fields producing fractal field reversal clusters shrinks as the ratio of interaction strengths becomes less degenerate, suggesting the experimental window for fractal behavior was narrow in previous studies~\cite{bingham2021experimental}. Pyrochlore systems lack this degeneracy, but their field reversal experiments constrained reversals to one dimension due to higher interaction energy at the surface as a result of fabrication~\cite{may2021magnetic}. Future experiments with finer field control and tuned interaction energies may reveal fractal field reversal clusters. However, the details of single island reversal are coarse-grained in this model. Real field reversal is facilitated by magnonic modes~\cite{zeissler2016low}, potentially altering the precise field reversal phase diagrams of experiments.

An implicit choice in constructing a nanomagnetic device is whether useful information is encoded in a low-energy state or the dynamic response of the system. The prevalence of energy-based logic gates for computation has diminished and dynamic responses for reservoir computing are on the rise~\cite{gartside2022reconfigurable,vidamour2022reservoir}. Directly driving a system overcomes some limitations of nanomagnets, allowing them to evolve at lower temperatures without freezing~\cite{morley2017vogel} and avoiding the critical slowing down of glassy systems stopping dynamics~\cite{souletie1985critical}. Real, disordered materials will nucleate field reversal clusters from relatively susceptible magnets and the complexity of the emerging structures will depend on field duration, direction, and strength. Initial studies on reservoir computing~\cite{jensen2020reservoir,hon2021numerical} already show specific windows of field and temperature in which the quality of the reservoir is high. Future analyses of systemic nanomagnet dynamics will further explore what ranges of parameters, system geometries, and the modes they facilitate are ideal for computation.

\begin{acknowledgments}
We thank Cristiano Nisoli and Will Branford for their insightful discussions. The work of F.C. and M.S. was carried out under the NNSA of the U.S., DoE at LANL, Contract No. DE-AC52-06NA25396 (LDRD Grant No. PRD20190195), LA-UR-23-21161.
\end{acknowledgments}

\appendix
\section{High Temperature Limit}
When temperature increases, the constituent spins in nanomagnets become paramagnetic. Mathematically, $\beta$ and its product with the interaction strengths and external field become much less than one, allowing for the approximation
\begin{equation}
    \dot{\vec{m}} = (\beta Q - \mathbb{1})\vec{m}+\beta \vec{h}(t).
\end{equation}

Recognizing this as an inhomogeneous system of linear differential equations, the equations admit a homogeneous and particular solution. The magnetization over time may be decomposed into a linear combination of eigenvectors of the matrix $B = \beta Q - \mathbb{1}$, $\mathbf{v}_k$, with corresponding eigenvalues $\lambda_k$, making the moments over time $\vec{m}(t) = \sum_k C_k(t) \vec{v}_k$. The homogenous solution is
\begin{equation}
    \vec{m}_h(t) = \sum_k C_k(0) \vec{v}_k e^{\lambda_k t}.
\end{equation}
For the ring specifically, the eigenvectors and eigenvalues for the $k$th mode are $v_{k,j} = \exp(i\frac{2\pi k}{N} j)$ and $\lambda_k = -1 + \beta q + 2\beta J\cos(2\pi k/N)$. All eigenvalues are negative since $\beta q << 1$ and $\beta J << 1$, meaning that all spatial modes that exist through initial conditions decay. The decay rate is slightly faster for low wavelength modes when $J$ is positive (ferromagnetism) and faster for long wavelength modes when $J$ is negative (antiferromagnetism). Ultimately, these homogeneous solutions vanish over time, removing the history of initial conditions and even the interaction structure of the magnets themselves.

The particular solution depends, as the name suggests, on the field protocol being applied. Considering a periodic field, it may be represented by Fourier series decomposition: $\vec{h}(t) = \sum_n \vec{H}_n \exp (i\omega_nt)$, where $\omega_n = 2\pi n/T$, with period $T$. The particular solution is then
\begin{equation}
    \vec{m}_p = \beta \sum_{n=1} \vec{H}_n\frac{1}{\omega_n} \exp \bigg(i \omega_n t - i\pi/2\bigg).
\end{equation}
The response leads the signal by a phase of $\pi/2$ and diminishes in magnitude proportional to the frequency, exactly the behavior of an RL circuit. Because this does not rely on the interaction matrix, we note that solutions to the mean-field Glauber equation at a high-temperature generally respond to external fields with uncoupled inductors like oscillations with exponentially fading hysteresis.

\section{Low-Temperature Solution on a Tree}
We begin with the differential equation
\begin{equation}
    \dot {\vec m}=-\vec m+\tanh\big(\beta(Q\vec m+\vec h)\big).
\end{equation}
We consider the limit $\beta\rightarrow \infty$, and we assume $Q_{ii}=0$ and $Q_{ij}=Q A_{ij}$, with $A_{ij}$ the adjacency matrix of the graph.
In this case, we have
\begin{equation}
    \dot {\vec m}=-\vec m+\text{sign}\big(Q A\vec m+\vec h\big).
\end{equation}
We now perform the change of variables $\vec m(t)=e^{-t} \vec r(t)$, and the differential equation becomes
\begin{equation}
    \dot {\vec r}= \text{sign}\big(e^{-t} Q A\vec r+\vec h\big).
\end{equation}
Since $A_{ii}=0$, the solution of the differential equation above can be written in the form
\begin{equation}
   \vec r(t)=\vec r(t_0)+ \int_{t_0}^t  \sinh\big(e^{-s} Q A\vec r(s)+\vec h\big)ds
\end{equation}
and thus
\begin{equation}
   \vec m(t)=e^{-(t-t_0)}\vec m(t_0)+e^{-t} \int_{t_0}^t e^s \text{sign}\big( Q A\vec m(s)+\vec h\big)ds
   \label{eq:soltree}
\end{equation}
We now make the following assumptions. First, we assume that $A_{ij}$ is the adjacency matrix of a loopless graph, e.g. a tree $\mathcal T$. Initially, the system is in the all up or all down states, e.g. $\vec m(t_0)=\pm 1$. Since the equation is invariant under the transformation $\vec m\rightarrow -\vec m$, $\vec h\rightarrow -\vec h$, we can choose an initial state, and obtain the solution for the, say, down state $-1$, and obtain the evolution for the other solution by flipping the sign of the solution and of the external field. If all initial spins are in the $-1$ state, a negative external field $h$ leaves the state invariant. 

We then assume that an initial spin $i$ is flipped $m_i(t_0)=+1$, and $h_i>0$. In the following, let us define $$F_i(t;a,b)\equiv-e^{-(t-t_0)} m_{i}(t)+e^{-t}\int_{t_0}^t e^s \text{sign}\big(Qa(s)+b(s)\big) ds.$$
\begin{figure*}
    \centering
    \includegraphics[width=0.75\textwidth]{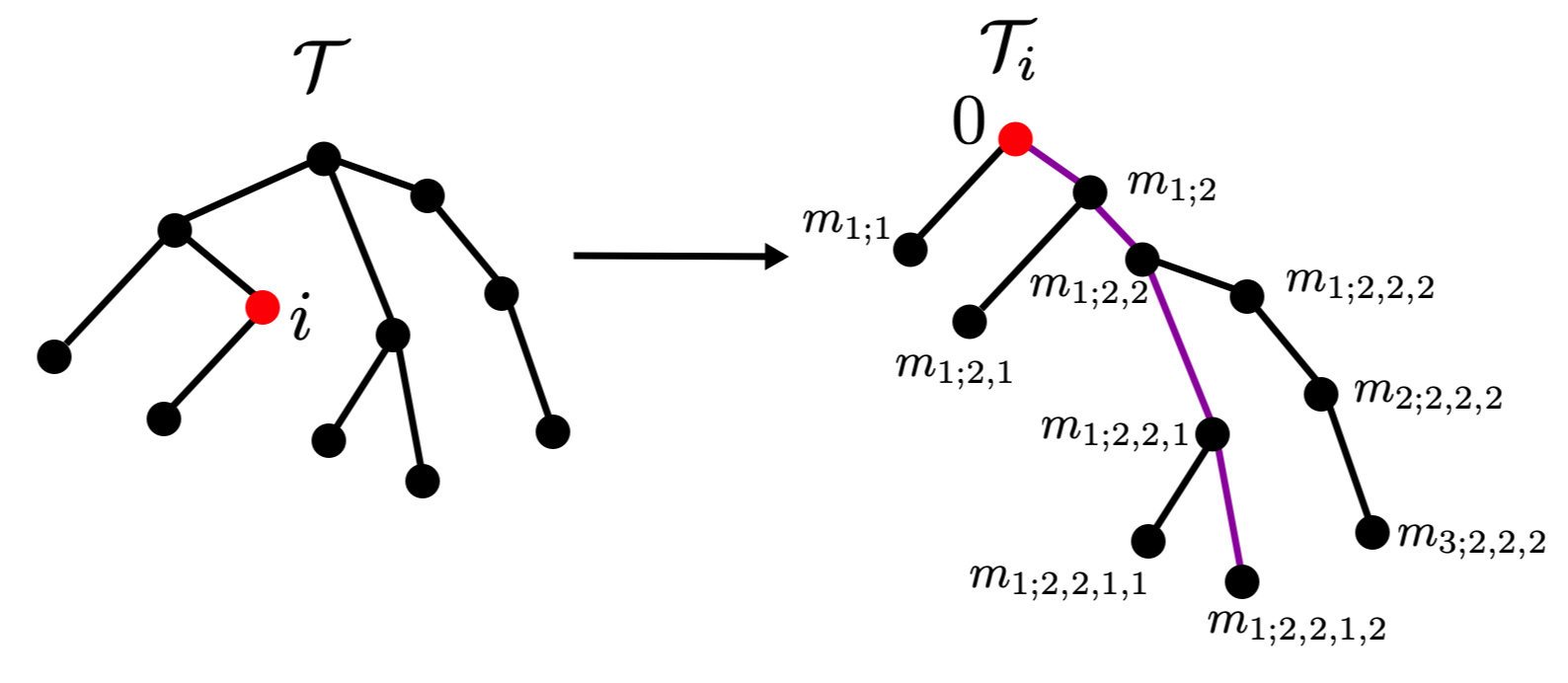}
    \caption{Tree $\mathcal T$ and its initial perturbation, and the rooted tree $\mathcal T_i$ at the perturbation. The end points of the tree are the children, and from the root to a child, one has to go through a branch.}
    \label{fig:rooted}
\end{figure*}
It is useful then to consider the tree $\mathcal T$ rooted at $i$. For a rooted tree, we set $i=0$ and assign the following coordinates to the spins in reference to the spin $0$.  For each child branch of $0$, we call the coordinates ${i_1,i_2,\cdots}$ where, $i^1_j$ is the $i^1$-th branch based at the root, ${i_1,i_2}$ represents the $i^2$ branch based at the branch $i^1$ and so on and so forth. Then $m_{k; i_1,i_2,\cdots}$ represents the $k$-spin on the branch ${i_1,i_2,\cdots}$. An example is shown in Fig.  \ref{fig:rooted}.

Since $m_0(t_0)=1$ and $h_0>0$, it is easy to see that $m_i(t)=\theta(t-t_0)$ from eqn. (\ref{eq:soltree}), if $h_0>Q$. This is the condition that perturbations are unstable. It is not hard to see that for $h>2Q$, all spins flip independently because the magnetic field.

We then have the chain of equations 
\begin{eqnarray}
m_0(t)&=&1\nonumber \\
m_{1;j}(t)&=& F_{1;j}(t;m_0+m_{2;j},h_{1;1}) \nonumber\\
&\vdots&\nonumber \\
m_{k_j;j}(t)&=&F_{1;j}(t;m_{k_j-1;j}+\sum_{r}m_{1;j,r},h_{k_j;j}) \label{eq:node}\\
&\vdots&\nonumber\\
m_{1;j,j^\prime}(t)&=&F_{1;j,j^\prime}(t;m_{k_j;j}+m_{2;j,j^\prime}(t),h_{1;j,j^\prime}) \nonumber\\
&\vdots&\nonumber
\end{eqnarray}
Since we are on a tree, what this shows is that we can consider any path from the root to the child, and these equations are decoupled subsets each topologically equivalent to the flipping of a spin on a chain graph, in which the spin at the beginning or the end of the chain is flipped at the initial state. Thus, solving it on the chain graph then provides the solution on each branch, as in the Lila subgraph in Fig.  \ref{fig:rooted}, provided that we properly change $h_i$ at the junction nodes, which we will discuss later. Let us first focus on a chain graph.
This is a chain of equations of the form :
\begin{eqnarray}
m_0(t)&=&1\nonumber \\
m_1(t)&=&F_1(t,m_0(t)+m_2(t),h_1)\nonumber\\%
m_2(t)&=&F_2(t,m_1(t)+m_3(t),h_2)\nonumber\\
m_3(t)&=&F_3(t,m_2(t)+m_4(t),h_3)\nonumber\\
\vdots
\end{eqnarray}
Let us now focus on the second equation. 
Since $m_i\geq -1$, we have the inequality
\begin{equation}
Q(m_{i-1}+m_{i+1})+h\geq Q(m_{i-1}-1)+h_i,
\label{eq:ineq}
\end{equation}
which we will use in a moment, since we assume $Q\geq0$.
    
Inside the sign function, we have $Q(1+m_2(t))+h_2\geq h_2$. Since, initially, $m_2(t_0)=-1$, and we have that $\text{sign}(h_1)=1$, it is easy to see that assuming $t_0=0$ the solution is given by,
\begin{eqnarray}
    m_1(t)=-\theta(-t)+\theta(t)(1-2 e^{-t}).
\end{eqnarray}
For $m_2(t)$, we have
\begin{eqnarray}
    m_2(t)&=&e^{-t} m_2(0)\nonumber \\
    &&+e^{-t}\int_{0}^t e^s \text{sign}\big(Q(1-2e^{-t}+m_3(t))+h_2\big) ds\nonumber 
\end{eqnarray}
and we have, because of the inequality (\ref{eq:ineq}), we have that
\begin{eqnarray}
    \text{sign}\big(Q(1-2e^{-t}+m_3(t))+h\big)\geq \text{sign}\big(Q(-2e^{-t})+h_2\big)\nonumber 
\end{eqnarray}
and thus, the sign switches when $-2Qe^{-t}+h=0$. We can define $\tau_2=\log(2Q/h)$, and after a little algebra, we find the solution
\begin{eqnarray}
    m_2(t)=-\theta(\tau_2-t)+\theta(t-\tau_2)\Big(1-2e^{-t+\tau_2}\Big). \nonumber
\end{eqnarray}
At this point, we see that for $m_3$ we have the same switching condition, but in which we replace $\tau_2$ with $\tau_3$, where 
\begin{eqnarray}
    \tau_3=\tau_2+\log(\frac{2Q}{h_3}).
\end{eqnarray}
The equation above then repeats at $m_4$, and we have thus the same set of solutions, defined by the recursive equations:
\begin{eqnarray}
    m_0(t)&=&1\\
    m_i(t)=&&-\theta(\tau_i-t)\nonumber \\
    &&+\theta(t-\tau_i)\Big(1-2 e^{-t+\tau_i}\Big)
 \end{eqnarray}
        where
    \begin{eqnarray}
    \tau_{i+1}&=&\tau_i+\log(\frac{2Q}{h_{i+1}})\\
    \tau_1&=&0.
\end{eqnarray}
or, if $h_i=h$, we have simply $\tau_{i}=(i-1)\log(\frac{2Q}{h}) $.
The speed of propagation of the signal is then given by 
\begin{eqnarray}
    v=\frac{\Delta x}{\Delta t}=\frac{1}{\log(\frac{2Q}{h})}.
\end{eqnarray}
The speed of propagation reaches infinity for $h=2Q$, which is the case in which every single spin is flipped independently from each other, only due to the effect of the external field. If instead the field is non-homogeneous, the effective speed is given by
\begin{eqnarray}
    \bar v=\frac{\Delta x}{\Delta t}=\frac{1}{\log(\frac{2Q}{ (\prod_{j=1}^n h_j)^{1/n}})}.
\end{eqnarray}
The difference between the chain and a tree is the presence of node equations, such as eqn. (\ref{eq:node}). These contain terms of the form
\begin{eqnarray}
    \text{sign}\big(Q(m_{k_{\vec j};\{\vec j\}}+\sum_r m_{1;\{\vec j,r\}})+h\big).
\end{eqnarray}
The equation above is an upper bound to
\begin{eqnarray}
    &&\text{sign}\big(Q(m_{k_{\vec j};\{\vec j\}}-d_{\{\vec j\}})+1+h\big)\nonumber \\
    &&=\text{sign}\big(Q(1-2 e^{-t +\tau_{k_{\vec j};\{\vec j\}}}-d_{\{\vec j\}}+1)+h\big)
\end{eqnarray}
Thus, the key difference between the chain and the tree, is that we need to replace 
\begin{eqnarray}
    h_{i}\rightarrow h_{i}-(d_i-2)Q
\end{eqnarray}
where $d_i$ is the node degree.
As a result, for the perturbation to propagate across nodes, we must have $h_{i}\geq (d_i-2)Q$ for all nodes. For a chain, $d_i=2$, and thus reduces to the chain solution above.




\bibliography{apssamp}

\end{document}